# Effects of thermal annealing and film thickness on the structural and optical properties of indium-tin-oxide thin films

Ding Xu[1], Wen Zhou[1,*], Yuxin Du[1], Junying Zhang[1], Wei Zhang[1,*], Jiangjing Wang[1,*]

[1]Center for Alloy Innovation and Design (CAID), State Key Laboratory for Mechanical Behavior of Materials, Xi'an Jiaotong University, Xi'an 710049, China.

E-mails: wen.zhou@xjtu.edu.cn, wzhang0@mail.xjtu.edu.cn, j.wang@xjtu.edu.cn

**Abstract:**

Indium-tin oxide (ITO) is a crucial functional layer for the optoelectronic applications, such as non-volatile color display thin films based on the ITO/phase-change material (PCM)/ITO/reflective metal multilayer structures on a silicon substrate. In addition to non-volatile color tuning by PCMs, thermally induced crystallization may alter the optical properties of ITO layers as well. But the potential change in color of the ITO layers is not considered so far. In this work, we investigate the structural and optical properties of ITO thin films via X-ray diffraction, spectroscopic ellipsometry and ultraviolet-visible spectrophotometry measurements. After thermal annealing at 250 °C, the ITO thin films of 15–100 nm get crystallized with strong changes in refractive index $n$ and extinction coefficient $k$ in the visible light range. However, for the 5-nm ITO thin film, crystallization is only observed after thermal annealing at 350 °C and the change in color is limited upon phase transition. We provide a colormap of the ITO/platinum/silicon structure in terms of the annealing temperature (150–350 °C) and ITO film thickness (5–100 nm). Our work suggests that the intrinsic change in colors of ITO layers should also be considered for the PCM-based reconfigurable display application.

## Introduction:

Indium-tin oxide (ITO) is a transparent semiconducting oxide with the advantages of electrical conductivity and high optical transmittance in the visible-light range[1-3]. ITO has various applications such as architectural coatings[4], gas sensors[5], humidity sensing[6], transparent electrodes in solar cells and flat panel displays[7-9]. Studies have been performed to control the microstructures and physical properties of ITO thin films through thermal annealing[10-12]. Chalcogenide phase-change materials (PCMs)[13-19] have been commercialized for high-density persistent memory and embedded memory technologies[20-25]. The basic working principle of phase-change memory devices is to utilize the fast and reversible phase transition between the amorphous phase and the crystalline phase of PCM, such as $Ge_2Sb_2Te_5$ (GST)[26-31], for memory programming. The large contrast in electrical or optical properties between the two solid-state phases is used to distinguish the basic logic states “0” and “1”. Upon phase transition, the reflectance of the PCM thin film also varies strongly, and has been used for rewritable optical data storage[32-35]. Recently, the ITO/phase-change material (PCM)/ITO/reflective metal multilayer deposited on silicon (Si) substrate[36] have been proposed for reconfigurable color display[37-45].

Figure 1a shows a schematic of a typical multilayer structure for display application. The thickness of the top ITO layer ($t_{ITO1}$) is typically 5–20 nm. It serves as a capping layer to prevent GST from oxidation. In contrast, the thickness of the bottom ITO layer ($t_{ITO2}$) is thicker (50–150 nm) to control the base color[36]. The reflective metal layer can be platinum (Pt). The structural phase transition of the GST layer from the amorphous to crystalline phases can be driven by thermal annealing on a hotplate or by applying weak but long optical or electrical pulses; the melt-quenching amorphization can be achieved using strong but short optical or electrical pulses. As shown in Figure 1b, crystallization of the GST is a multifold phase transition, including amorphous to cubic phase transition under ~150–210 °C annealing and cubic to hexagonal phase transition under ~250–350 °C annealing[46] through a vacancy ordering process[47-54]. These transitions can produce significant changes in optical properties for tunable color display.

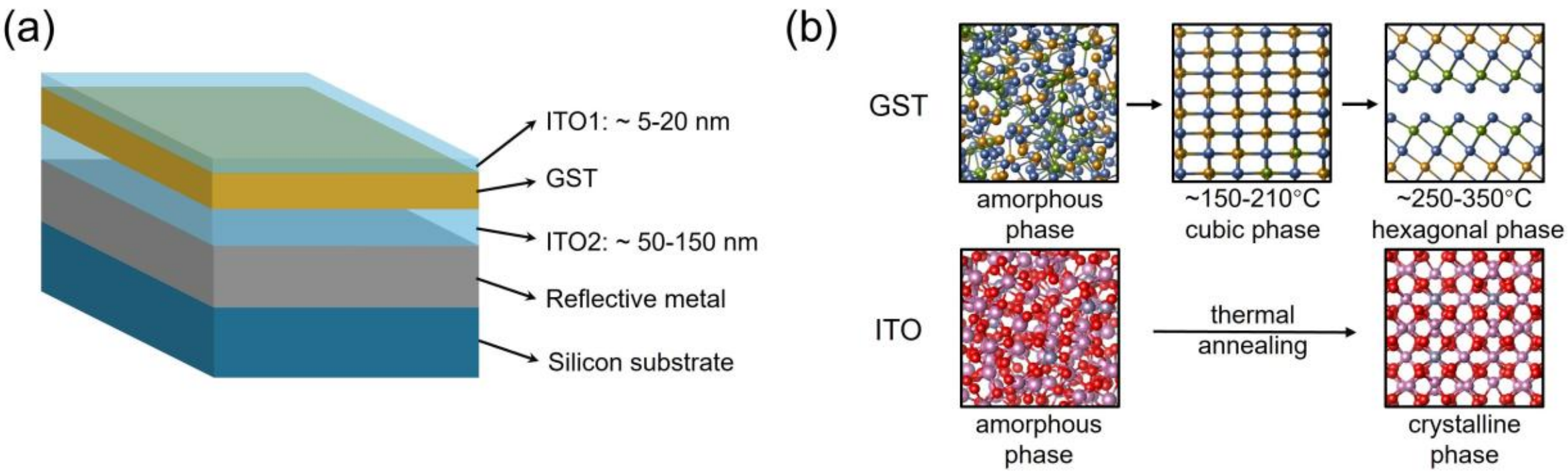

**Figure 1.** (a) Schematic of a reflective display multilayer thin-film structure consisting of ITO/GST/ITO/reflective metal layers on a Si substrate. (b) Phase transition of GST and ITO by thermal annealing.

As reported in Ref.[55], the 200-nm amorphous ITO thin film shows a crystallization temperature of ~250 °C, and the crystalline ITO thin film would display much different optical properties as compared to the amorphous phase[55]. Hence, when the amorphous PCMs with a higher crystallization (≥ 250 °C), the intrinsic changes in color of ITO should also be considered. For example, $CrTe_3$ (~270 °C)[56], $CrGeTe_3$ (~270 °C)[57-60] and related two-dimensional layered materials[61-63], Ge-rich GST (above 350 °C)[64-66] and carbon-doped GST (330 °C)[67] all show a higher crystallization temperature than that of amorphous ITO. Also, it is known that the thermal stability of amorphous materials can be largely enhanced as the film thickness is reduced to a few nm. Hence, it is important to provide a color map as function of thermal annealing temperature and ITO film thickness. In this work, we investigate the crystallization and optical properties of amorphous ITO thin films of 5–100 nm on standard silicon substrates before and after thermal annealing. The changes in structure and optical profiles are studied via X-ray diffraction and spectroscopic ellipsometry experiments. Then we conduct ultraviolet-visible spectrophotometry measurements to record the color changes of various ITO/Pt/Si multilayers upon thermal annealing.

## Results

Using magnetron sputtering, we deposited ITO thin films of various thickness to study the structural and optical properties of the thin films upon thermal annealing induced phase transition. If the sputtering target contains multiple elements, the obtained as-deposited thin films usually form an amorphous phase. Figure 2a shows a schematic of an ITO thin film deposited on a Si substrate (ITO/Si) with a film thickness of $t_{ITO}$, and Figure 2b shows the optical images of ITO thin films with different thicknesses (5 nm, 15 nm, 30 nm, 50 nm, and 100 nm). Clearly, the color of the ITO/Si thin film is altered by tuning the thickness of the ITO layer, which stems from the optical interference effect[36]. Figure 2c shows a cross-sectional scanning electron microscopic (SEM) image of the ITO sample, in which the thickness of ITO thin film was measured as ~100 nm. According to our cross-sectional SEM characterizations (Fig. S1), the measured thickness of the deposited and annealed ITO films match with the target thickness (5–100 nm). The variations in film thickness before and after thermal annealing are less than 1 nm. All the deposited and annealed ITO films exhibit smooth and flat surfaces according to the top-view SEM images (Fig. S2). Figure 2d shows the energy-dispersive X-ray spectroscopy (EDS) measurement results for the ITO thin films. It exhibits characteristic peaks corresponding to indium (In), tin (Sn), Si, and oxygen (O) elements. Our analysis reveals the relative atomic percentage ratio of In:Sn to be approximately 9.9:1, which matches well with the atomic ratio (9.8:1) of our ITO target ($In_2O_3$/$SnO_2$ 90/10 wt%).

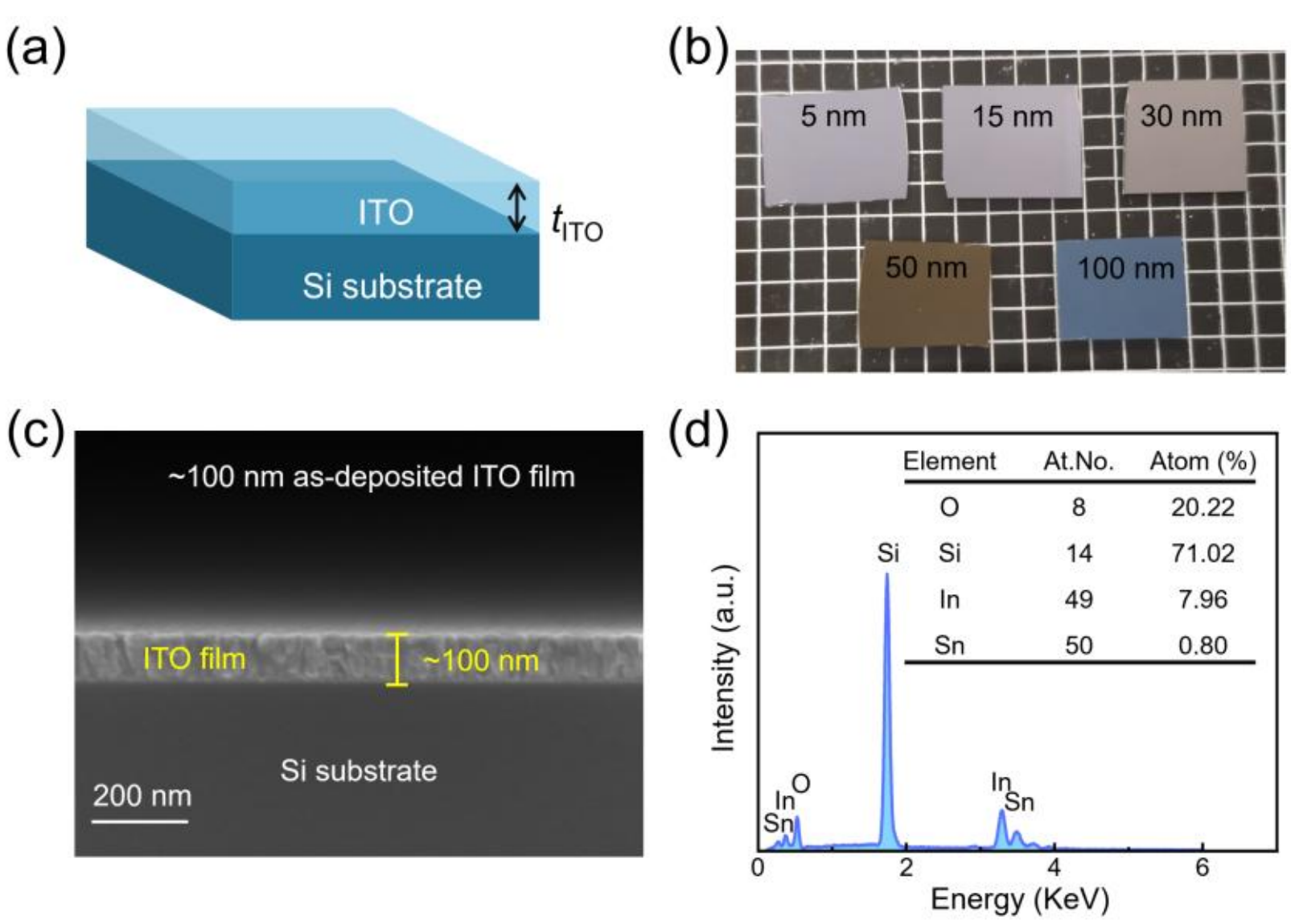

**Figure 2.** Characterizations of the ITO/Si samples. (a) Schematic of an ITO layer on a Si substrate. (b) Photograph of ITO thin films deposited on Si substrates with varied ITO thickness ($t_{ITO}$). (c) A cross-sectional SEM image of an ITO/Si sample. (d) The EDS analysis of an ITO thin film.

Figure 3 shows the X-ray diffraction (XRD) patterns of ITO samples with different thicknesses before and after annealing at temperatures of 150 °C, 200 °C, 250 °C and 350 °C, respectively. Before annealing, all the as-deposited ITO thin films displayed no clear diffraction peaks except the Si (211) peak (Fig. 3a), indicating that all the as-deposited ITO films formed an amorphous phase. It should be noted that the major Si (211) peak was not always observed in the XRD patterns. This is because we used single-crystal Si substrate, but the crystallographic orientation of the substrate did not always match the Bragg condition. We slightly rotated the sample specimen, the same Si (211) peak was consistently observed (Fig. S1). All the ITO films annealed at 150 °C show no characteristic peaks of ITO, because the annealing temperature was still well below the crystallization temperature. After thermal annealing at 250 °C, 15-nm ITO film showed a diffraction peak at 30.74°, corresponding to the (222) peak of crystalline ITO. The (211), (400) and (440) peaks appeared in thicker ITO thin films, indicating that these ITO thin films were crystallized. Regarding the 5-nm ITO film, the (222) diffraction peak only emerged after thermal annealing at 350 °C (Fig. 3b). Although the intensities of XRD peaks weakened in thinner films, the appearance of the (222) peak was sufficient to confirm the crystallization of the ITO thin film. It is noted that ITO was already crystallized in the 100-nm thin film after annealing at 250 °C, but crystallization only occurred in the 5-nm thin film after annealing at 350 °C. This enhanced amorphous stability in the ultrathin films of a few nm in thickness can be attributed to the nano-confinement and volume effects [68-70]. For example, monatomic Sb is known to crystallize spontaneously at room temperature, but when the thickness is scaled down to a few nm, the lifetime of amorphous Sb can be largely increased [71-75]. In the 2-nm case, the crystallization temperature of amorphous Sb can even reach 243 °C[76].

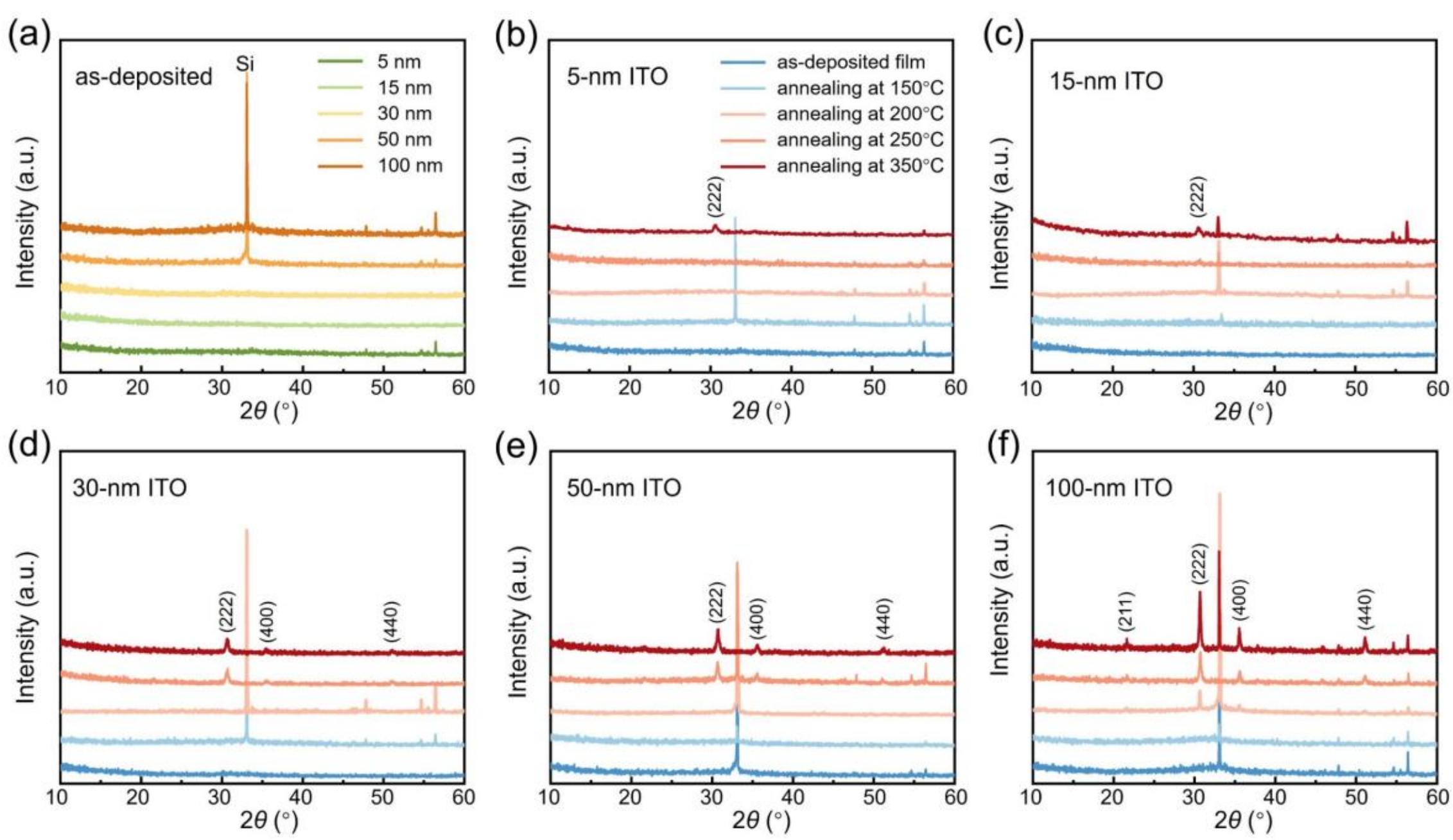

**Figure 3.** XRD patterns of the ITO/Si samples. (a) XRD patterns of the as-deposited ITO with thicknesses of 5–100 nm. (b–f) XRD patterns of the as-deposited and annealed ITO with thicknesses of 5 nm (b), 15 nm (c), 30 nm (d), 50 nm (e), and 100 nm (f). These ITO/Si samples were annealed at 150 °C, 200 °C, 250 °C, and 350 °C, respectively.

We next investigated the thickness-dependent optical properties of the ITO films. The ITO samples discussed above were used for spectroscopic ellipsometry experiments. The optical constant of the thin films in a wavelength range of 200–2100 nm was modeled using the CODE software, and the detailed fitting process can be found in the supplementary note 1 and Table S1. Figure 4a shows the measured optical constants (*n*, *k*) of the as-deposited ITO samples with thicknesses of 5–100 nm. In the visible light regime (400–760 nm), the refractive index (*n*) and extinction coefficient (*k*) of the as-deposited ITO films increase gradually with increase of ITO film thickness. Subsequently, the ITO films were annealed at various temperatures (150–350 °C). Obvious changes in the optical constants occur upon crystallization at 350 °C for the 5-nm ITO film and 250 °C for the 15-nm to 100-nm ITO films as shown in Fig. 4b–f, which are consistent with the estimated crystallization temperatures by the XRD analyses. The annealed and crystallized ITO films show decrease in refractive indices and increase in extinction coefficients[77]. This can be explained by increase in the free carrier concentration after crystallization, which leads to enhancement in light absorption. Moreover, according to the Drude model describing the plasmonic behavior, increase in free electron density introduces a redshift in the plasma wavelength, which results in a decrease in the refractive index near the plasma edge[78]. These results consistently demonstrate that the phase transition introduces obvious changes in the optical constants of ITO films.

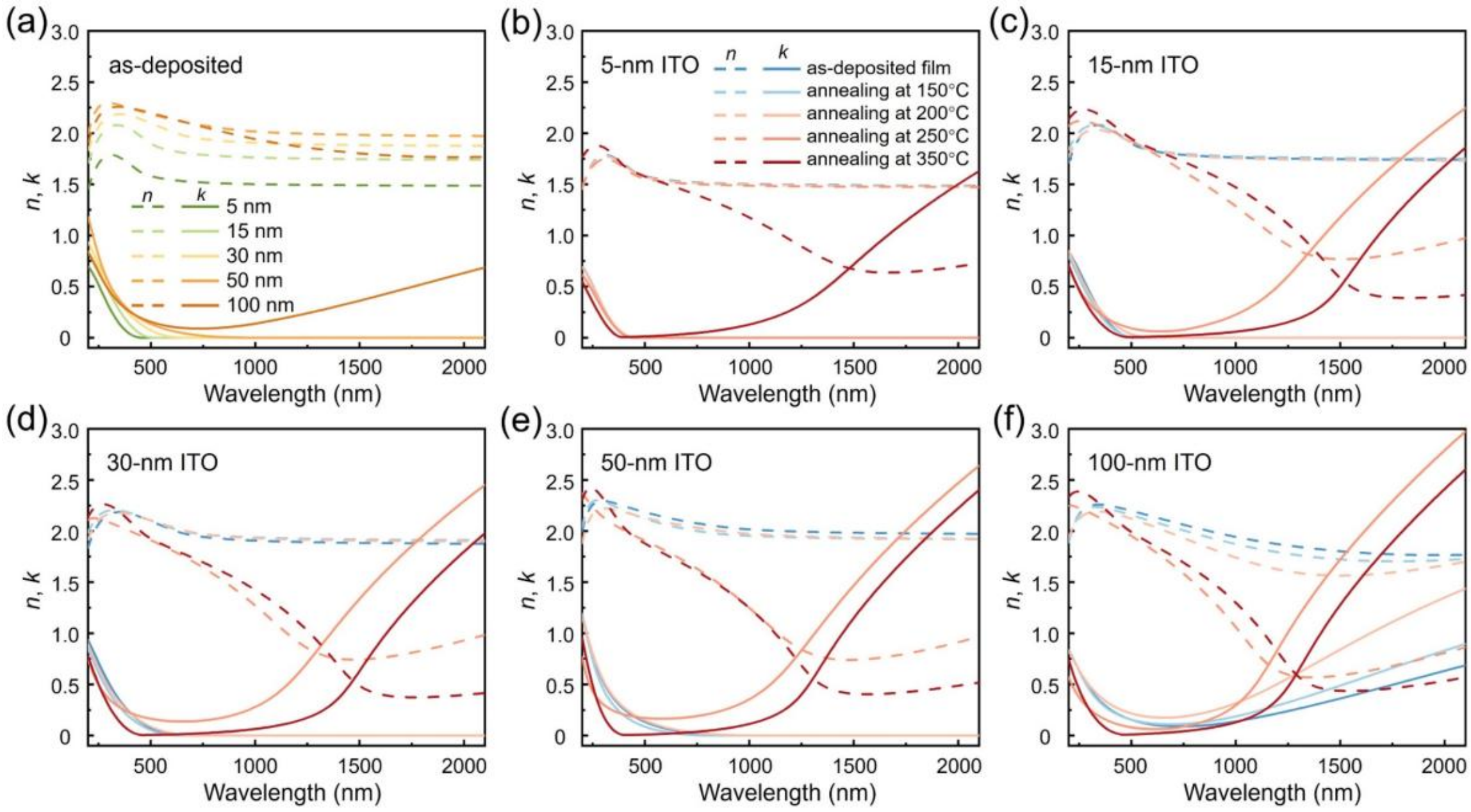

**Figure 4.** Spectroscopic ellipsometry measurements of the ITO/Si samples. (a) Ellipsometry measurement of refractive indices of the as-deposited ITO with thicknesses of 5–100 nm. (b–f) Measured refractive indices of the as-deposited and annealed ITO with thicknesses of 5 nm (b), 15 nm (c), 30 nm (d), 50 nm (e), and 100 nm (f). These ITO/Si samples were annealed at 150 °C, 200 °C, 250 °C, and 350 °C, respectively.

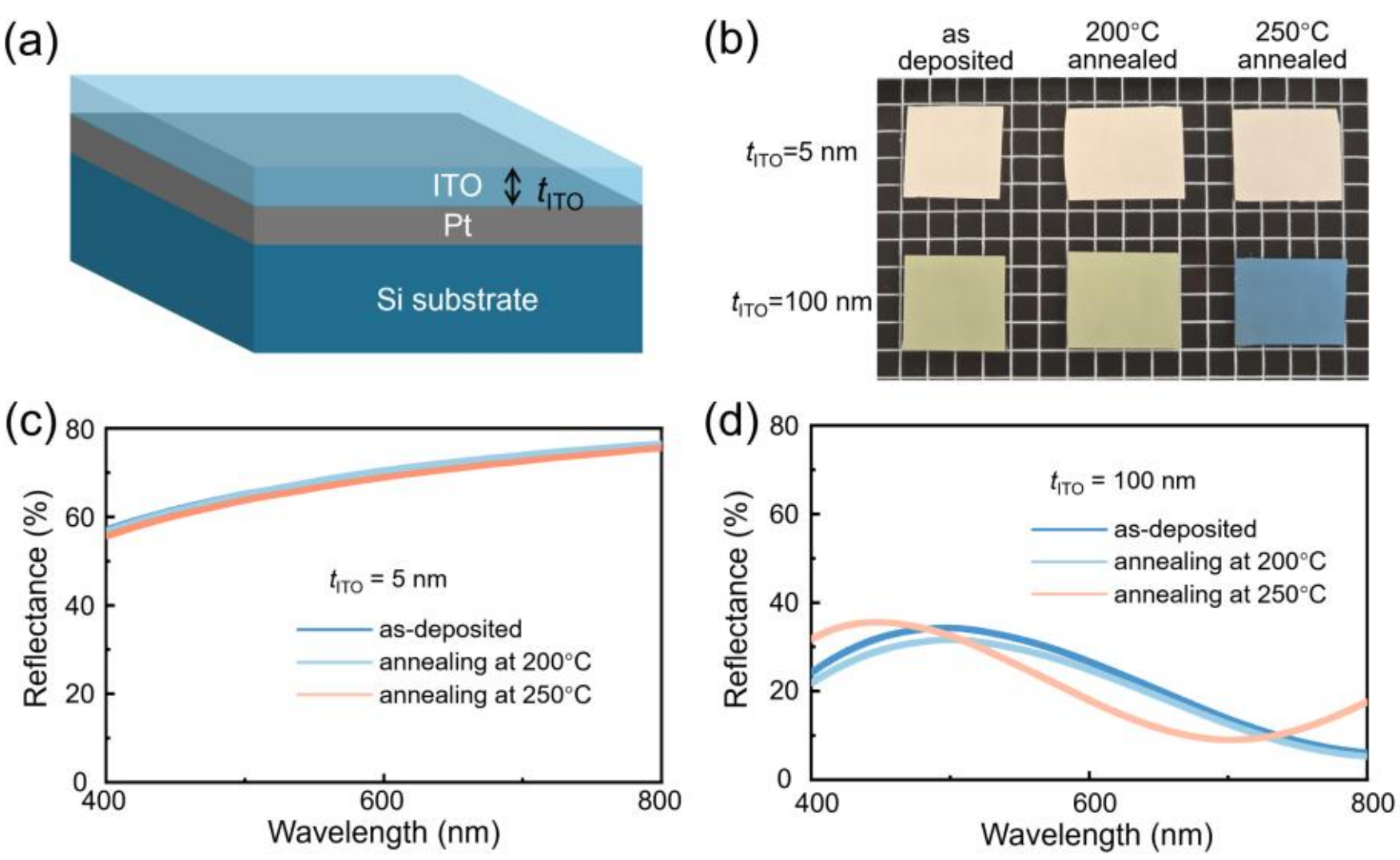

**Figure 5.** Structure and properties of the ITO/Pt/Si samples. (a) Schematic of a reflective display thin-film structure consisting of an ITO layer and a Pt mirror coated on a Si substrate. (b) As-deposited and annealed samples with ITO thicknesses of 5 nm and 100 nm. (c) and (d) Reflection spectra of the as-deposited and annealed samples with ITO thicknesses of 5 nm (c) and 100 nm (d).

To investigate color regulation of ITO, we proposed a basic thin-film structure consisting of an ITO layer with a thickness of $t_{ITO}$ and a Pt mirror deposited on a Si substrate (ITO/Pt/Si) with a schematic as shown in Fig. 5a. These thin films can act as a resonant optical cavity, enabling tailorable colors by adjusting the thickness or phase state of ITO. The ITO samples were further annealed at 200 °C and 250 °C to investigate tunable reflective colors. As shown in Fig. 5b, when the thickness of ITO increases from 5 nm to 100 nm, the reflective color changes from light pink to yellow for the as-deposited ITO samples. The observed color changes in the samples clearly indicate that ITO thickness plays a crucial role in defining base colors. The colors of 5-nm and 100-nm-thick ITO samples show no notable change after thermal annealing at 200 °C. However, a notable color change from yellow to blue was observed in the 100-nm-thick ITO sample annealed at 250 °C due to crystallization. Figure 5c and Figure 5d show reflection spectra of the as-deposited and annealed samples with ITO thicknesses of 5 nm and 100 nm, respectively, measured by ultraviolet-visible spectrophotometry. For the 5-nm-thick ITO samples, it shows almost no variation in the reflection spectra after thermal annealing because the crystallization gets more difficult due to the nano-confinement and volume effects [68-70]. However, the 100-nm-thick ITO samples can already be crystallized at 250 °C with a strong modulation in refractive index, and thereby there is an obvious blue shifting of the resonant peak wavelength from 500 nm to 450 nm by increasing annealing temperature from 200 °C to 250 °C. This spectral blue shifting is consistent with the observed color evolution of the samples from yellow to blue as shown in Fig. 5b.

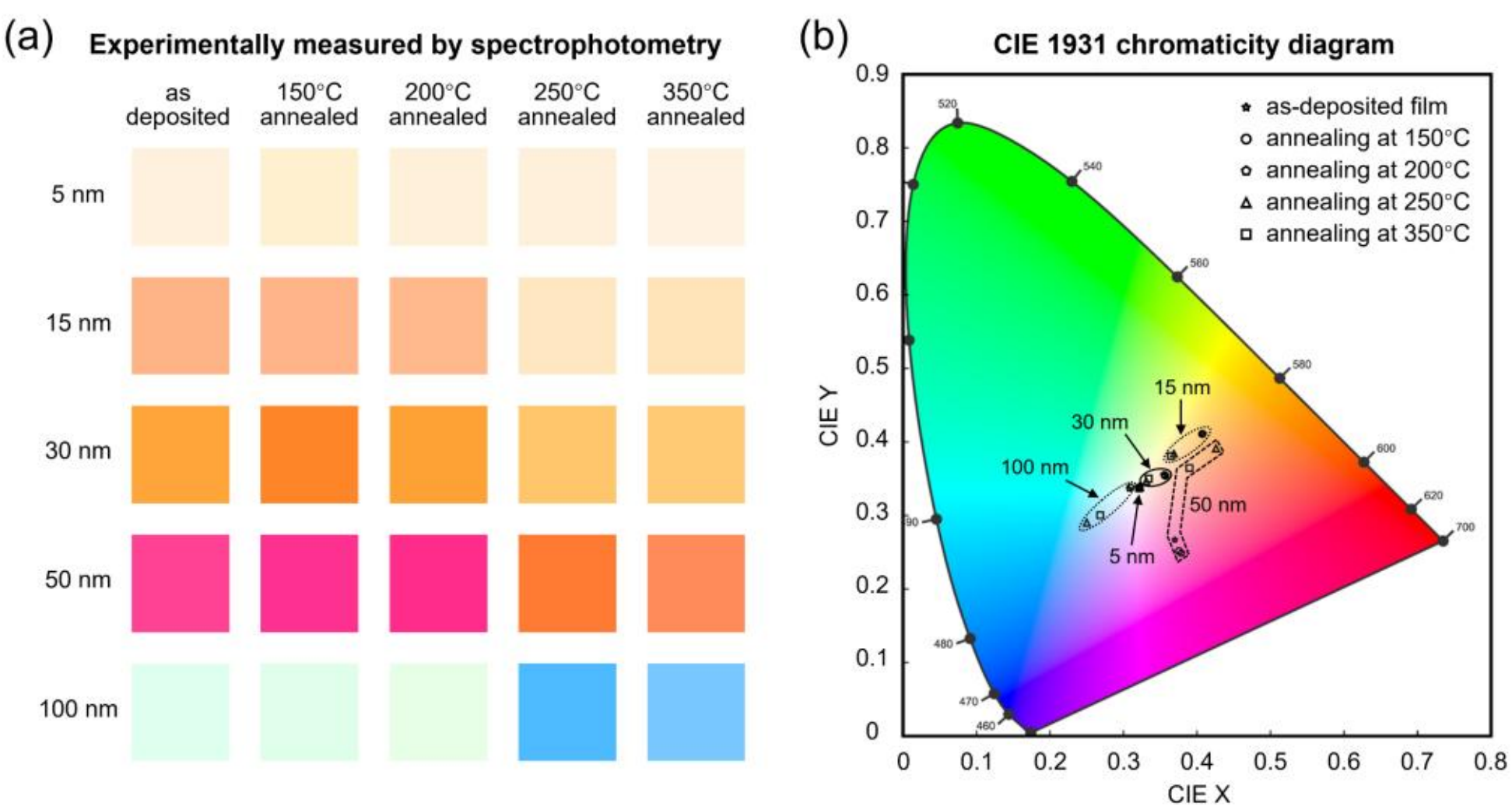

**Figure 6.** Colormap of the ITO/Pt/Si samples. (a) Tunable reflecting color of the ITO/Pt/Si samples with varied ITO thickness (5–100 nm) and thermal annealing temperature (150–350°C). (b) CIE1931 chromaticity diagram showing the achievable reflective colors of the as-deposited and annealed ITO samples with varied thickness.

To systematically examine color tunability of the ITO samples, we experimentally measured the reflection spectra of the samples with varied ITO thickness and annealing temperatures by ultraviolet-visible spectrophotometry. We next converted the reflection spectra into the red-green-blue (RGB) colors to elaborate color display of these ITO thin films as shown in Fig. 6a. It shows that the reflecting color can be tunable by increasing the ITO thickness or thermal annealing above the crystallization temperatures. As shown in Fig. 6a, the chromaticity difference between the annealed and as-deposited ITO becomes more pronounced by increasing thickness. Notably, it shows obvious color variation under 250 °C annealing treatment for the as-deposited ITO samples with thicknesses of 50 nm and 100 nm. We further mapped the reflection spectra onto the CIE1931 chromaticity diagram by using the CIE color-matching equation[79]. These achievable colors marked on the CIE1931 diagram are clearly separated, except for those of the 5-nm-thick ITO samples as shown in Fig. 6b. Among them, we can observe a relatively large color tuning range for the 50-nm-thick ITO samples before and after crystallization. Thus, the ITO/Pt/Si multilayer structures can be exploited for color display with obvious color changes when the ITO thickness is around or above 50 nm and annealing temperature above 250 °C. Phase transition behavior of the ITO thin films may be useful for optical steganography and encryption applications[80].

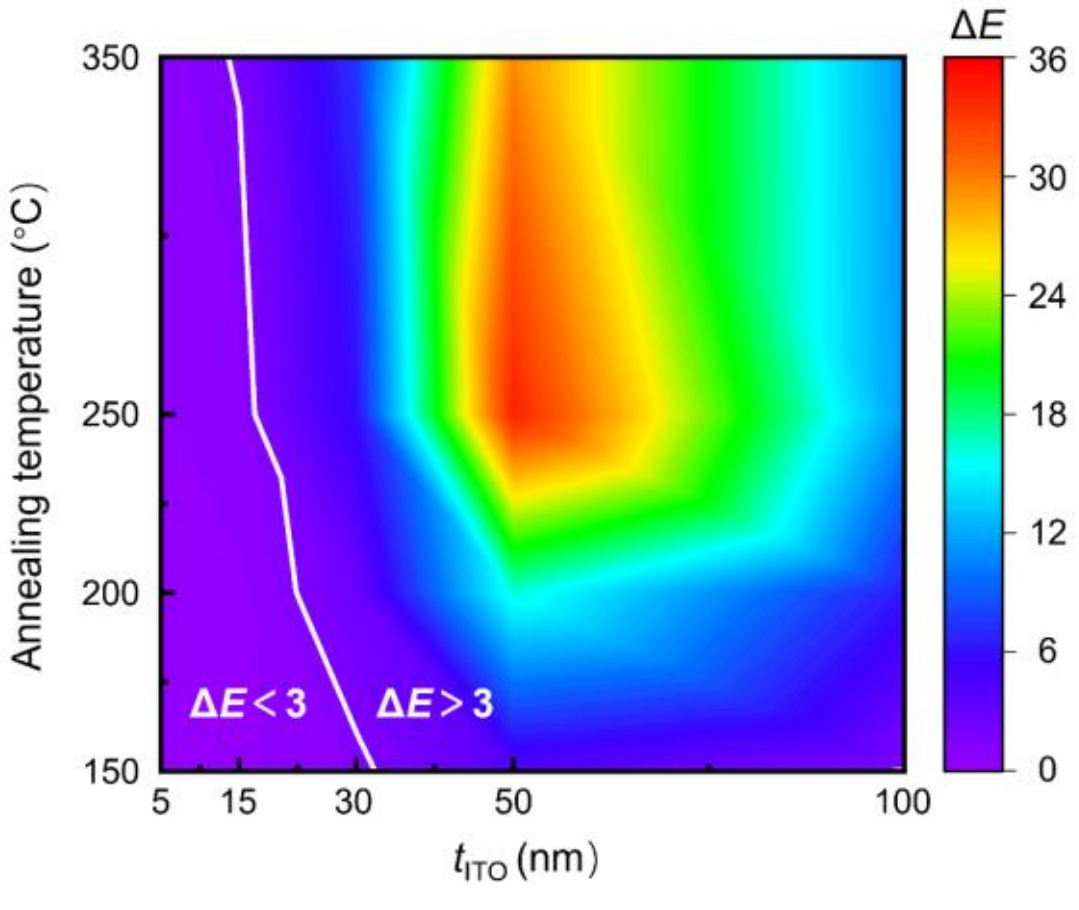

**Figure 7.** The contour of color difference ($\Delta E$) of the ITO/Pt/Si samples with varied ITO thickness and annealing temperature.

To quantitatively examine the color modulation performance, the color difference ($\Delta E$) between the as-deposited and annealed ITO/Pt/Si samples was calculated using the CIEDE2000 formula recommended by the International Commission on Illumination (CIE)[81]. This formula provides superior visual perception agreement compared to the CIE76 standard[82]. Figure 7 shows a calculated contour of $\Delta E$ with varied ITO thickness and annealing temperature. Based on the threshold of human eyes for color perception, we defined $\Delta E$ = 3 as the evaluation criterion, and color changes are considered visually perceptible when $\Delta E$ exceeds this value. In particular, if crystallization occurs, the $\Delta E$ value is sharply increased, e.g., from 5.72 to 35.02, when the annealing temperature is to increased 250 °C (ITO thickness = 50 nm). In contrast, ITO layers impose a very weak effect in color modulation when the thickness is less than 15 nm, regardless of the thermal annealing temperature (150–350 °C).

## Summary

In summary, we systematically investigated effects of thermal annealing and film thickness on the structural and optical properties of ITO thin films. The 5-nm ITO thin film showed crystallization temperature of ~350 °C, much higher than that of the thicker films. The refractive indices of ITO thin films with varied thicknesses on silicon substrates were characterized by ellipsometry experiments, which showed strong changes upon crystallization. Although the changes in $n$ and $k$ were strong for the ITO/Si structure with $t_{ITO}$=5 nm after crystallization, the color of the ITO/Pt/Si multilayer structure remained nearly the same after thermal annealing at 150–350 °C. For thicker ITO films with $t_{ITO}$=30, 50 and 100 nm, crystallization induced a major change in refractive indices of the ITO/Si structure as well as in the color of the ITO/Pt/Si multilayer structure. The largest change in color was observed for the ITO/Pt/Si multilayer structure with $t_{ITO}$=50 nm. To reduce the impact of ITO-induced color changes in the ITO/PCM/ITO/Pt/Si multilayer for practical display applications, we suggest to keep the thickness of top ITO layer to be 5 nm but to use thicker bottom ITO layer with $t_{ITO} \geq$ 100 nm.

## Methods

Thin-film deposition: The ITO target ($In_2O_3$/$SnO_2$ 90/10 wt%) with a purity of 99.99% was used for magnetron sputtering. ITO thin films were deposited on silicon wafers using the magnetron sputtering system (AJA Orion-8). Deposition process was carried out with a base pressure at $\sim 1\times10^{-7}$ Torr and an argon atmosphere at 3 mTorr. At room temperature, ITO thin films with thicknesses ranging from 5 nm to 100 nm were prepared by adjusting the deposition time with a direct current (DC) power of 60 W. The Pt metal mirrors were fabricated by depositing a 5-nm-thick Cr adhesion layer and a 100-nm-thick Pt layer via DC sputtering at 50 W power onto a silicon substrate. The deposition rate of ITO, Cr and Pt layers was 1.8 nm/min, 4.2 nm/min, 7.1 nm/min, respectively. Annealing of the as-deposited ITO films were performed in an argon atmosphere using an Instec mK200 hot stage. The ITO samples were heated to the targeted temperature in the range of 150 °C to 350 °C at a ramp rate of 10 °C/min, were held at each target temperature for 10 min, and were subsequently cooled down to room temperature.

Material characterization: compositions of the thin-film samples were characterized using a scanning electron microscope (SEM; Sigma 300, Zeiss) with build-in energy-dispersive spectroscopy (EDS;

XFlash 7, Bruker). The thicknesses of the deposited ITO thin films were directly measured by the cross-sectional SEM. X-ray diffraction (XRD) measurements were performed using a Bruker D8 ADVANCE diffractometer with Cu-Kα radiation ($\lambda$ = 1.54056 Å) over a 2$\theta$ range of 10°–60° and a scanning step of 0.02°. Reflectance spectra were measured via ultraviolet-visible spectrophotometry (Lambda 950, PerkinElmer) with a 2-nm step resolution across a wavelength range of 400–800 nm. Spectroscopic ellipsometry measurements were performed using a rotating compensator ellipsometer (UVISEL PLUS, Horiba). The optical constant was obtained by fitting the experimental data using the CODE software (https://wtheiss.com). The ITO/Si sample was modeled as a two-layer structure. The dielectric function of the ITO layer was described by a combination of the Tauc-Lorentz, Drude oscillator models and a constant background term.

**Acknowledgments**
The work is supported by 111 Project (B25007). The authors acknowledge the Instrumental Analysis Center of Xi'an Jiaotong University.

**Conflict of interest**
The authors declare no conflict of interest.

**Author Contributions**
**Ding Xu**: data curation (lead); formal analysis (lead); investigation (lead); writing—original draft (equal). **Wen Zhou**: conceptualization (equal); data curation (equal); investigation (equal); writing—original draft (equal). **Yuxin Du**: data curation (equal); investigation (equal). **Junying Zhang**: data curation (equal); investigation (equal). **Wei Zhang**: conceptualization (lead); funding acquisition (lead); supervision (lead); writing—review and editing (lead). **Jiangjing Wang**: conceptualization (equal); supervision (equal); writing—review and editing (equal).

**Data Availability Statement**
The data that support the findings of this study are available from the corresponding author upon reasonable request.